# Unexpected edge conduction in mercury telluride quantum wells under broken time reversal symmetry

Eric Yue Ma[1,2], M. Reyes Calvo[1,2], Jing Wang[1,3], Biao Lian[1,3], Mathias Mühlbauer[1,5], Christoph Brüne[5], Yong-Tao Cui[1], Keji Lai[1,2,4], Worasom Kundhikanjana[1,2], Yongliang Yang[1], Matthias Baenninger[1,3], Markus König[1,3], Christopher Ames[5], Hartmut Buhmann[5], Philipp Leubner[5], Laurens W. Molenkamp[5], Shou-Cheng Zhang[1,3], David Goldhaber-Gordon[1,3], Michael A. Kelly[1], & Zhi-Xun Shen[1,2,3*]

[1]Geballe Laboratory for Advanced Materials, Stanford University, Stanford, California 94305, USA

[2]Department of Applied Physics, Stanford University, Stanford, California 94305, USA

[3]Department of Physics, Stanford University, Stanford, California 94305, USA

[4]Department of Physics, University of Texas at Austin, Austin, Texas 78712, USA

[5]Physikalisches Institut (EP3), Universität Würzburg, Am Hubland, D-97074, Würzburg, Germany

*email: zxshen@stanford.edu

# Abstract

The realization of quantum spin Hall effect in HgTe quantum wells is considered a milestone in the discovery of topological insulators. Quantum spin Hall states are predicted to allow current flow at the edges of an insulating bulk, as demonstrated in various experiments. A key prediction yet to be experimentally verified is the breakdown of the edge conduction under broken time-reversal symmetry. Here we first establish a systematic framework for the magnetic field dependence of electrostatically gated quantum spin Hall devices. We then study edge conduction of an inverted quantum well device under broken time-reversal-symmetry using microwave impedance microscopy, and compare our findings to a non-inverted device. At zero magnetic field, only the inverted device shows clear edge conduction in its local conductivity profile, consistent with theory. Surprisingly, the edge conduction persists up to 9 T with little change. This indicates physics beyond simple quantum spin Hall model, including material-specific properties and possibly many-body effects.

# Introduction

Quantum spin Hall (QSH) theory predicts that when the thickness of an HgTe quantum well (QW) exceeds a critical value $d_C$, a band inversion happens and the QW transitions from a trivial insulator to a QSH insulator with gapless edge channels protected by time-reversal symmetry (TRS)[1,2]. Whereas edge conduction in the presence of TRS has been well established in various transport[3–7] and microscopy studies[8,9], its predicted breakdown under broken TRS remains unverified[10,11], except for earlier works showing suppression of conductance by weak field in devices with strong backscattering at zero field[3], and a gap opening at the otherwise protected Landau level (LL) crossing near the field-induced topological phase transition[12,13]. A systematic magnetic-field-dependent study of QSH edge conduction combining microscopy and transport techniques is therefore highly desirable to gain further insight into this key prediction.

In this work we combine transport and microwave impedance microscopy (MIM), a scanning probe technique that measures local electromagnetic response[14–16], to study two HgTe/($Hg_{0.3}Cd_{0.7}$)Te QW devices, above (7.5 nm) and below (5.5 nm) the theoretical $d_C$ of 6.7 nm. At zero magnetic field, edge conduction is only observed in the 7.5 nm device, consistent with QSH theory. In the case of broken TRS by a magnetic field, we first develop a rigorous treatment for the field dependence of electrostatically gated HgTe QW devices and show that the widely accepted picture of re-entrant behavior in Hall conductance[3] needs to be revisited. We then discuss our field-dependent MIM and transport data. They agree well with our expectations in general, but with one surprise: the edge conduction in the 7.5 nm device survives up to 9 T with no sign of disappearing, well beyond the expected crossover field of ~3.8 T, when the QW should transition into a trivial insulator. The stark contrast suggests the necessity of including more physics to understand QSH in real systems. We discuss the possibility of conduction through trivial edge states enhanced by spin-orbit coupling and effects of electron-electron interactions.

# Results

## 5.5 vs. 7.5 nm quantum well in zero magnetic field

We first establish the influence of the different QW thicknesses in transport measurements. We use a two-step etching process to define mesas with clean physical edges for both devices (Fig. 1a,

Supplementary Fig. 1 and Supplementary Note 1). The conductive substrate (*n*++ GaAs) 4 μm below the QW is used as the back gate to tune the QW from *p*-type through the bulk gap to *n*-type, as shown in the 2-terminal resistance ($R_2$) taken at 4.5 K and 0 T, of a strip 50 μm long by 1225 μm wide (Fig. 1c, f). When the Fermi level ($E_F$) is gate-tuned to the bulk gap, $R_2$ reaches 6 MΩ (150 MΩ/□) for the 5.5 nm QW but only 50 kΩ for the 7.5 nm QW, consistent with contribution from edge conduction.

We then use microwave impedance microscopy (MIM) to confirm the edge conduction in real space. MIM delivers a small excitation (<~0.1 μW) at 1 GHz to a metallic tip and measures the reflected signal with a spatial resolution of ~150 nm. The two output channels, MIM-Im and MIM-Re, are proportional to the change of the imaginary and real parts of the tip-sample admittance[15]. For a buried QW structure, the MIM-Im signal is an intuitive semi-logarithmic representation of QW conductivity with a calibrated two-tone color scale (Fig. 1b, Supplementary Fig. 2 and Supplementary Note 2). Repeated MIM single-line scans are performed across a ~5 μm wide mesa while tuning the gate voltage, producing Fig. 1d (5.5 nm) and 1g (7.5 nm). When $E_F$ is in valence or conduction band, both the 5.5 and 7.5 nm device appear homogeneously conductive ($\sigma_{xx}$ >~2x$10^{-6}$ $\Omega^{-1}$); when $E_F$ is tuned toward the bulk gap, the 5.5 nm device becomes homogeneously insulating ($\sigma_{xx}$ <~2x$10^{-8}$ $\Omega^{-1}$) while the 7.5 nm device only becomes insulating in the bulk - the edge remains conductive (Fig. 1d, g, e, h, also Supplementary Fig. 3).

The smooth MIM profile of the edge conduction with a full-width-half-maximum (FWHM) of ~1.5 μm (green trace in Fig. 1h) is not limited by the spatial resolution. Instead it reveals that local conductivity gradually decreases by ~3 orders of magnitude going from the edge to the center of the mesa. Whereas the FWHM of local conductivity in linear scale (<~500 nm) is consistent with previous results[7], the origin of the small MIM signal extending ~2 μm into the bulk requires better understanding[17].

## Expected magnetic field dependence for electrostatically gated HgTe quantum well devices

We now move on to the magnetic field dependence. Fig. 2a and 2f are calculated LL fan charts for 5.5 and 7.5 nm QW. We included an inversion symmetry breaking term to reproduce the observed avoided crossing of the 'zero modes' at the cross-over field of ~3.8 T (Fig. 2f)[12,13,18]. Previously the LL fan charts were directly compared with gate- and field-dependent transport data to demonstrate a 're-entrant quantum Hall effect' in an inverted device: $E_F$ was thought to stay constant with fixed gate voltage and thus if resides in the gap at zero field, will cross $\nu$ = +(-)1 LL and back to the gap as the field increases[3]. This picture needs to be revisited, because in electrostatically gated devices, a constant gate voltage corresponds to a constant *carrier density* instead of constant $E_F$; the excess carriers needed for $E_F$ to cross the lowest LLs when starting from the gap are simply not supplied (Supplementary Note 3).

Fig. 2b and 2g are the calculated *bulk* density of states (DOS) from Fig. 2a and 2f with vertical axis converted to carrier density using a disorder-broadened LL DOS (Methods): they can now be directly compared to the gate- and field-dependent transport and MIM scans. Note that Fig. 2b and 2g are very similar despite drastic differences between 2a and 2f, because the DOS of a LL is independent of the LL fan chart and only depends on $B$ (Supplementary Fig. 4). The difference in field dependence between inverted and non-inverted QW devices is therefore much more subtle than previously understood, as demonstrated below.

## Magnetic field dependence of 5.5 nm quantum well

The field dependence of the 5.5 nm device agrees well with expectation (Fig. 2a-e). Gate- and field-dependent 2-terminal resistance (Fig. 2d) shows a simple metal-insulator transition (MIT) at low densities, with the insulating region expanding linearly with field. MIM scans of a 7.2 μm long section of the mesa confirm that the mesa stays mostly homogeneously conductive or insulating with smooth transitions (Fig. 2e). The observed behavior can be well understood by assuming that the mobility edge is at a fixed LL filling factor $\nu$ (~+0.15/-0.2 in this particular device, orange and blue lines in Fig. 2c-e).

## Magnetic field dependence of 7.5 nm quantum well

The field dependence of the 7.5 nm device can be best illustrated through comparison with the 5.5 nm device. We start with expectations from a simple QSH model. Comparing Fig. 2g and 2b, we know the bulk should behave similarly in both devices, insulating between the two mobility edges ($|\nu| < \sim 0.2$) and conductive outside this range. The major predicted difference lies in the regime with insulating bulk (Fig. 2h): edge states are expected to exist between 0 and 3.8 T (Region I) due to the anomalous bending of the 'zero modes' near physical edges (Fig. 2f inset), but become progressively weaker due to disorder scattering and evolution of the 'zero modes', until completely disappearing into the bulk at 3.8 T, at which point the QW becomes a 'trivial insulator' (Region II)[19]. We stress here that the LL filling factor in region I and II is low enough ($|\nu| < \sim 0.2$) that no integer quantum Hall effect should be present, as opposed to previous expectations[3] (Supplementary Fig. 5). The gradual disappearance of the edge conduction from region I to II should be directly observable in real-space images; a reduced resistance should also be measured in region I due to the edge conduction, contrasting with a fully insulating behavior in region II.

In the measured 2-terminal resistance data on the 7.5 nm device, however, the expected fully-insulating region II breaks into two sub-regions II and II* (Fig. 2i), where II* is not insulating. The nature of this discrepancy is readily revealed in the MIM images (Fig. 2j): whereas the bulk has already become insulating in II*, edge conduction, which is predicted to disappear beyond the cross-over field (3.5±0.5 T as determined experimentally), persists up to 9 T with no sign of disappearing, therefore making the QW continue to appear conductive in transport. The edge conduction gradually disappears towards the *p*-type side; the QW thus becomes homogeneously insulating and the resistance goes up by two orders of magnitude in region II of Fig. 2i. The boundary between II and II* (green dashed line in Fig. 2i) has a similar slope as the bulk mobility edge ($\nu \sim +0.2$, orange dashed line) but with an offset in density, and therefore extrapolates to a different 'charge-neutral point' at 0 T (green dotted line). The regions where edge conduction is present (Region I and II*), although artificially divided, show virtually no discontinuity. In fact, the edge conduction appears even clearer at finite field (compare 0 T/1.0 V and 5 T/1.0 V in Fig. 2j), as the bulk becomes more homogeneously insulating, likely due to the localization caused by the magnetic field.

A careful analysis of the MIM line profile further demonstrates that the local edge conductivity dominates electronic transport in both low- and high-field regimes. Fig. 3a and 3c show the averaged line-cuts of the MIM images in the two columns corresponding to the 3 T and 7 T case in Fig. 2j. The extracted edge and bulk MIM signal is plotted against $V$g in Fig. 3b, d, together with the corresponding $R_2$ data. It is immediately obvious that both the peak position and magnitude of $R_2$ have a strong correlation with the edge signal, but not the bulk signal. This excellent agreement between local signal and transport confirms that the unexpected edge conductivity is due to edge states that can conduct DC current across mesoscopic distances, instead of localized states at the edges.

## Discussion

These results in HgTe QWs represent an opportunity to learn physics beyond the simple textbook QSH theory – a rare occurrence in the field of topological insulators where so far theoretical predictions are routinely confirmed. One might naively attribute the observed edge conduction in moderate to high fields to trivial edge states, for example due to a unidirectional band bending (Fermi level pinning) or fringing field effect from gating[20]. Such trivial edge states, if exist, are often assumed to be localized and not contributing to transport[21–23]. But in this particular material, strong spin-orbit coupling may suppress direct backscattering[24] and thus "protect" the conduction via trivial edge states, possibly to mesoscopic lengths. In fact, both scenarios can account for what we see going from the bulk gap to the *n*-type side: edge conduction turning on before bulk conduction. However neither scenario can account for our observations near the *p*-type side: Fermi level pinning would predict that the edge remains insulating when the bulk becomes conductive; fringing field effect would predict edge conduction turned on before the bulk becomes conductive;

whereas experimentally the conductivity of edge and bulk evolve similarly upon entering the *p*-type side (e.g. $V$g = 0 to -1.5 V in Fig. 3d).

Moreover, the edge conduction, expected and unexpected, only appears in the 7.5 nm device – suggesting that it cannot be completely 'trivial', but is related to the thickness induced band inversion. Additional crucial ingredients may include many-body effects[25–28] and material specifics of HgTe QW, including its strong spin-orbit coupling, *n*-*p* asymmetry, mercury vacancies[21,29,30] and intrinsic surface effect[22,23]. As it stands now, our discovery could herald broadly-relevant physics, or simply reflect specifics of this material. In either case, with HgTe QWs being the standard model system for QSH, our results should shed light on extending existing theory to treat other potential contributions and understanding QSH in real systems.

# Methods

## Quantum well growth and device fabrication

Each of the two fully strained HgTe/($Hg_{0.3}Cd_{0.7}$)Te quantum wells was grown by molecular beam epitaxy (MBE) on a 4 μm fully relaxed CdTe buffer atop an *n*++ GaAs substrate. The mobility exceeds 200,000 $cm^2$ $V^{-1}$ $s^{-1}$ in the *n*-type regime of the 7.5 nm device. QW thickness was determined by X-ray reflectometry and magnetotransport measurements. The QW is buried 50 (26) nm under the top surface in the 7.5 (5.5) nm device. The long-range surface topographic features mostly come from the relaxation of the CdTe layer instead of the QW or the capping layer. Gate-QW capacitance was extracted from Hall measurement and was used to convert gate voltage to carrier density and LL filling factor. A two-step etching process was used to create mesas ~130 nm deep with clean physical edges for transport and MIM scans (Supplementary Fig. 1 and Supplementary Note 1).

## MIM measurement and response curve calculation

MIM measurements were implemented using an Attocube-based scanner stack in a Janis He4 cryostat with 9 T superconducting magnet[15]. Topography was taken with a quartz tuning fork (TF) sensor to which the MIM probe is attached; MIM images were taken separately with the tip ~30 nm above mesa surface in constant height mode. The efficient capacitive coupling at microwave frequency enables sub-surface sensing without inducing interband transitions. The tip-sample interaction is in the near-field limit, so the spatial resolution is determined by the tip diameter (~150 nm) rather than the microwave wavelength. Imaginary and real part of the complex tip-sample admittance were recorded during scanning to produce the MIM-Im and -Re channel images. Only $\sigma_{xx}$ contributes to the screening of the microwave electric field; $\sigma_{xy}$ is not probed here. A small bias is applied to the tip to compensate the work function difference. The relationship between QW conductivity and tip-sample admittance was obtained using COMSOL Multiphysics 4.4 (details see Supplementary Note 2).

## Band structure and LL fan chart calculation

The critical thickness of 6.7 nm was determined by an 8-bulk-band model for a fully strained HgTe/($Hg_{0.3}Cd_{0.7}$)Te QW. We employ a 4-2D-band effective $k \cdot p$ Hamiltonian for HgTe quantum well to calculate the LL fan chart:

$$H = \begin{pmatrix} h(\mathbf{k}) & 0 \\ 0 & h^*(-\mathbf{k}) \end{pmatrix} + H_{\mathrm{BIA}} + H_{\mathrm{Z}} \ .$$

The kinetic term is a 2×2 matrix

$$h(\mathbf{k}) = \epsilon(\mathbf{k})I + \begin{pmatrix} m(\mathbf{k}) & A(k_x + ik_y) \\ A(k_x - ik_y) & -m(\mathbf{k}) \end{pmatrix},$$

where an expansion up to $k^4$ is used:

$$\epsilon(\mathbf{k}) = C + Dk^2 + \frac{F_4 k^4}{1 + F_6 k^6}\ ,\ m(\mathbf{k}) = M + Bk^2 + \frac{G_4 k^4}{1 + G_6 k^6}\ ,$$

with the coefficients determined from first principle calculations. The inversion symmetry breaking BIA term and Zeeman term are:

$$H_{\mathrm{BIA}} = \begin{pmatrix} 0 & 0 & 0 & -\Delta \\ 0 & 0 & \Delta & 0 \\ 0 & \Delta & 0 & 0 \\ -\Delta & 0 & 0 & 0 \end{pmatrix}, \quad H_{\mathrm{Z}} = \mu_{\mathrm{B}} B \begin{pmatrix} g_{\mathrm{E}} & 0 & 0 & 0 \\ 0 & g_{\mathrm{H}} & 0 & 0 \\ 0 & 0 & -g_{\mathrm{E}} & 0 \\ 0 & 0 & 0 & -g_{\mathrm{H}} \end{pmatrix},$$

where $\Delta = 1.5{\sim}2\mathrm{meV}$ depending on the quantum well thickness. In a magnetic field $B$, we calculate the Landau levels fan chart by substituting $\mathbf{k} \to \mathbf{k} - e\mathbf{A}$ . In the presence of the BIA term, there is an anti-crossing of Landau levels with a gap $2\Delta$ in an inverted sample. Calculated LL fan charts are converted to DOS intensity plots using the disorder-broadened LL DOS of

$$\mathrm{DOS}(E, E_{\mathrm{LL}}) = \frac{eB}{h}\frac{1}{\pi}\frac{\Gamma/2}{(E - E_{\mathrm{LL}})^2 + (\Gamma/2)^2}$$

with $\Gamma = 0.5$ meV.

## Acknowledgements

We thank Chao-Xing Liu for helpful discussions. This work was supported by Defense Advanced Research Projects Agency Microsystems Technology Office, MesoDynamic Architecture Program (MESO) through the contract number N66001-11-1-4105 to support E.Y.M. The work at Stanford was also supported by Center for Probing the Nanoscale, an NSF NSEC under grant PHY-0830228 to Z. X. S. and D. G.-G. to partially support Y.L.Y, NSF grant DMR1305731 to partially support W.K. and Y.T.C, the Gordon and Betty Moore Foundation through Grant GBMF3133 and Foundation's EPiQS Initiative through Grant GBMF4546 to Z. X. S. to partially support Y.L.Y and Y.T.C., the European Union under the project FP7-PEOPLE-2010-274769 to M. R. C., and the Department of Energy, Office of Basic Energy Sciences, Division of Materials Sciences and Engineering, under contract DE-AC02-76SF00515 to S. C. Z. to support J. W. for theoretical calculations and to D. G.-G. for device fabrication and characterization. The work in Würzburg was also supported by the German Research Foundation (DFG grant HA5893/4-1 within SPP 1666 and the Leibniz Program), the EU ERC-AG program (Project 3-TOP) and the Elitenetzwerk Bayern program "Topologische Isolatoren".

## Author contributions

E. Y. Ma did MIM and transport measurements. M. R. Calvo and M. Mühlbauer developed the fabrication process and fabricated the devices. C. Ames, P. Leubner and M. R. Calvo characterized the material. C. Ames and P. Leubner grew the QW wafers. J. Wang and B. Lian did the theoretical calculations. M. Baenninger fabricated devices used for preliminary measurements. Data analysis, interpretation and manuscript preparation were done with all authors.

## Competing Financial Interests

M.A.K and Z.X.S are co-founders of PrimeNano Inc. which licensed the sMIM technology from Stanford for commercial instrument.

# Figures Legends

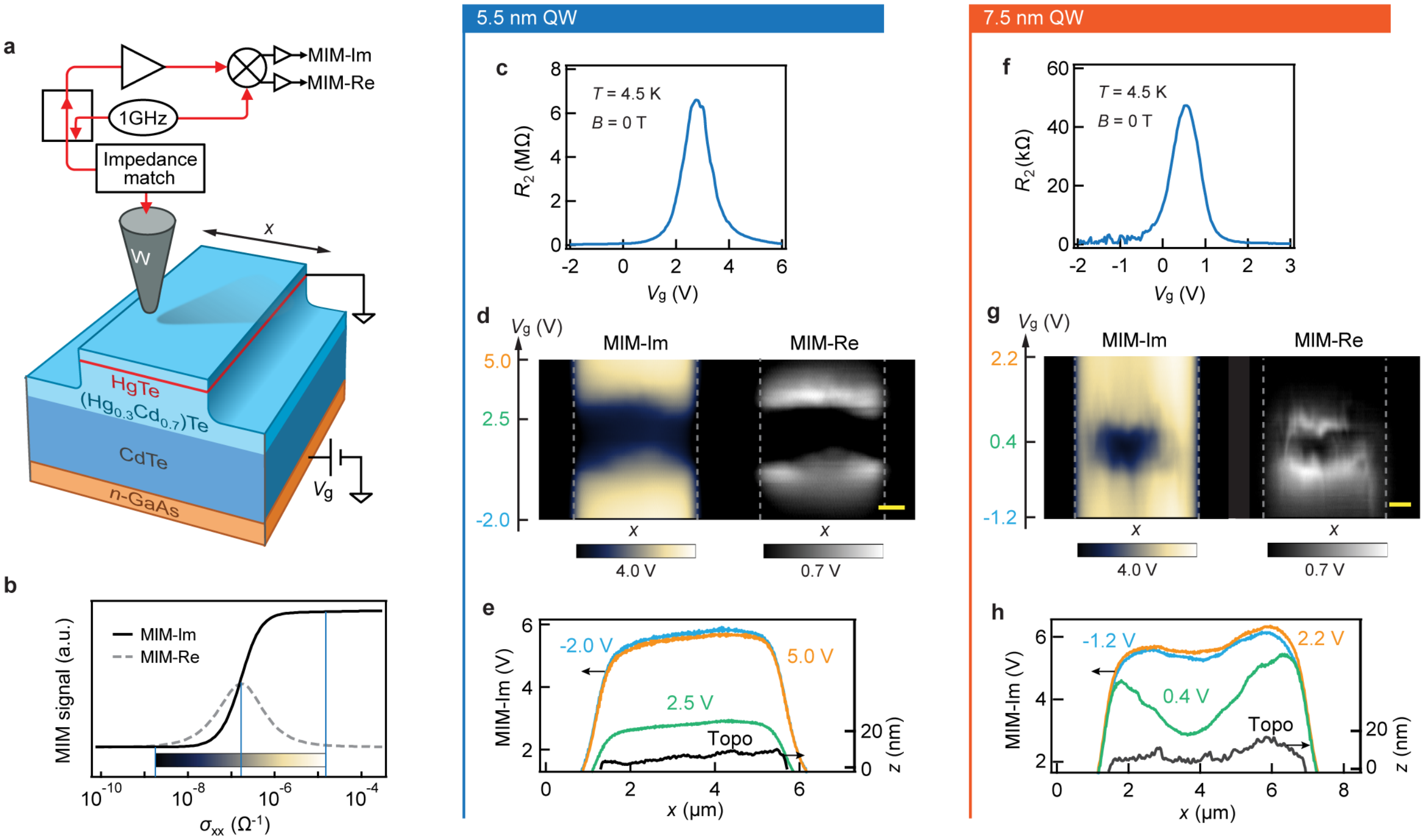

**Figure 1. Microwave impedance microscopy (MIM) setup and zero field gate dependence of 5.5 and 7.5 nm devices. (a)** Schematic of MIM setup and device structure, showing line scan direction *x*. **(b)** MIM response as a function of the sheet conductance of the quantum well, obtained by finite element analysis. The sensitivity window is centered at $2x10^{-7}$ $\Omega^{-1}$ and extends roughly two orders of magnitude in both directions, as indicated by the two-tone color scale. **(c)** Gate dependent 2-terminal resistance of a 50 μm long stripe on the 5.5 nm device. **(d)** Typical MIM-Im and MIM-Re single-line scan vs. gate voltage ($Vg$) across a ~5 μm wide mesa at 4.5 K. Physical edges are indicated by grey dashed lines. Two-tone color scale is used in MIM-Im to reflect the MIM sensitivity window in **b**, with neutral grey corresponding to the maximum in MIM-Re. The mesa appears homogeneously insulating with no edge conduction in the bulk gapped region. The scale

bar is 1 μm, for *x* direction only. **(e)** MIM-Im line-cuts when the QW is in *p*-type (-2.0 V), gap (2.5 V) and *n*-type (5.0 V). Black solid line is the mesa surface topography detail. MIM stays in constant-height mode with the tip ~30 nm above the average mesa surface; when moving into the etched-away region outside the mesa, the tip-sample distance quickly increases to ~160 nm and thus the MIM-Im signal decreases significantly. For the same reason the right part of the mesa has a slightly higher MIM-Im signal due to higher topography. This crosstalk only contributes a gate-independent modulation to the MIM signal. **(f-h)** Same measurements for the 7.5 nm device. Note the lower maximum 2-terminal resistance and prominent edge conduction in the gapped region. Crosstalk from topography is again visible as a gate-independent modulation of MIM signal.

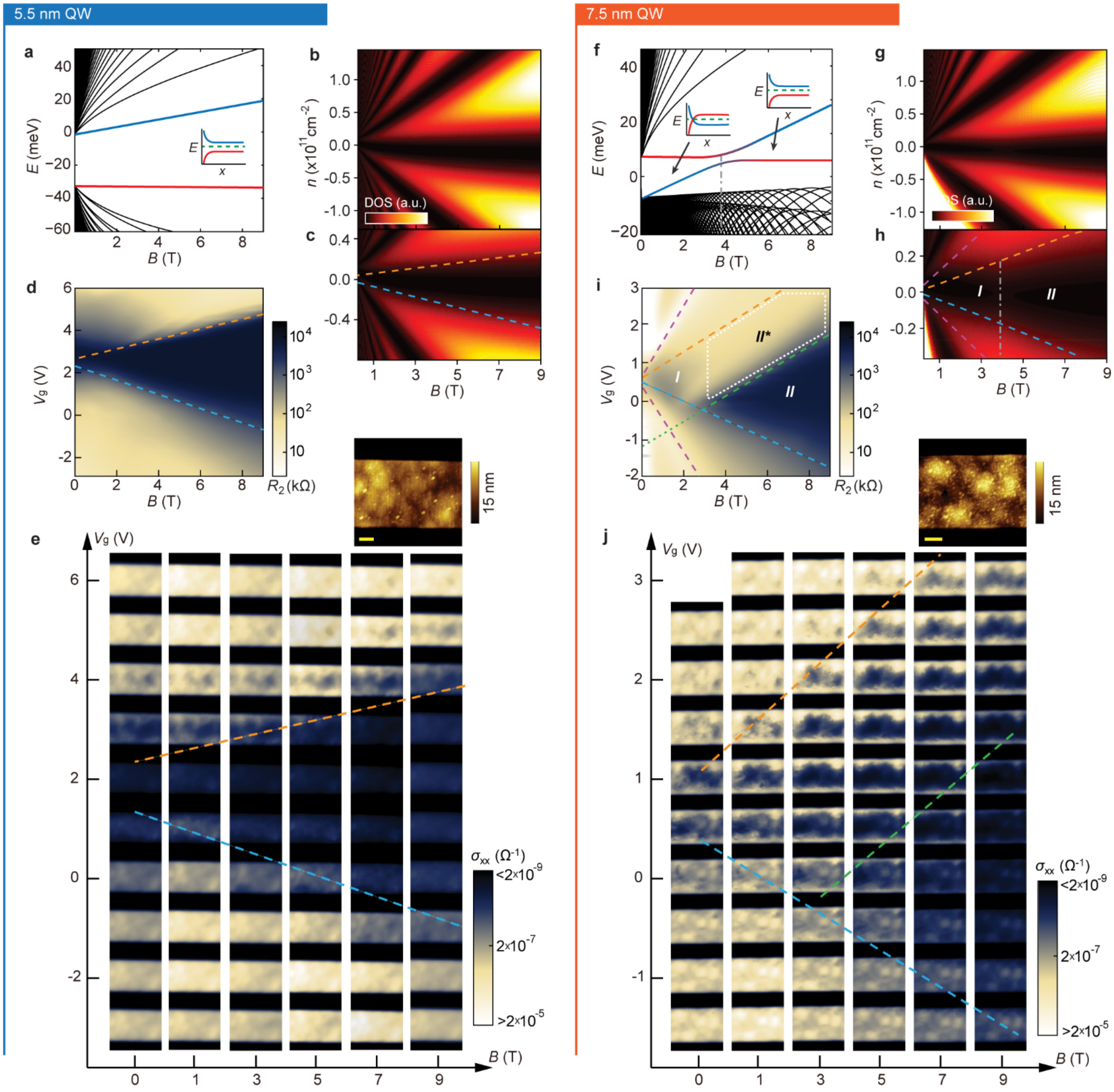

**Figure 2. Magnetic field dependence of the 5.5 and 7.5 nm devices. (a)** Calculated LL fan charts of the 5.5 nm device. Inset shows the bending of the electron-like (blue) and hole-like (red) lowest LLs near a physical edge. Green dashed line is the Fermi level. **(b)** Bulk DOS vs. carrier density and magnetic field, converted from **a**. The conversion is necessary because the back gate linearly tunes

the carrier density in the QW rather than the relative position between $E_F$ and the LLs. **(c)** Zoomed in of **b** corresponding to the density range achievable with our back gate. Electron and hole side mobility edges are indicated. **(d)** 2-terminal resistance vs. back gate voltage and magnetic field at 4.5 K. For ease of comparison, the color scale is chosen such that the corresponding sheet resistance range corresponds to the MIM sensitivity window in zero field limit ($\sigma_{xy} = 0$). **(e)** Corresponding real space MIM-Im images. The two-tone color scale reflects the MIM sensitivity window. Inset shows the surface topography details so one can identify topography crosstalk. The scale bar is 1 μm. The linearly broadened insulating regime in **d** and **e** corresponds to mobility edges at fixed filling factors (orange and blue dashed lines, as marked in **c**). **(f-j)** Same calculations and experimental results for the 7.5 nm device. Note the anomalous bending of the 'zero modes' in **f** inset. Because the estimated disorder broadening (~0.5 meV) and thermal broadening (~0.35 meV) are much smaller than the anti-crossing gap (~4 meV), the bulk is expected to remain insulating at the cross-over field of 3.8 T as shown in **h** (Supplementary Fig. 4). Note the smaller tuning range in the 7.5 nm device. The major unexpected feature is the absence of insulating behavior in region II* in **i**, which corresponds to edge conduction that survives up to 9 T as seen in **j**. The 'trivial insulator' phase only appears near the *p*-type side. Purple dashed lines in **i** correspond to $\nu \sim \pm 0.5$ (also marked in **h**), visible in the 7.5 nm device due to higher mobility.

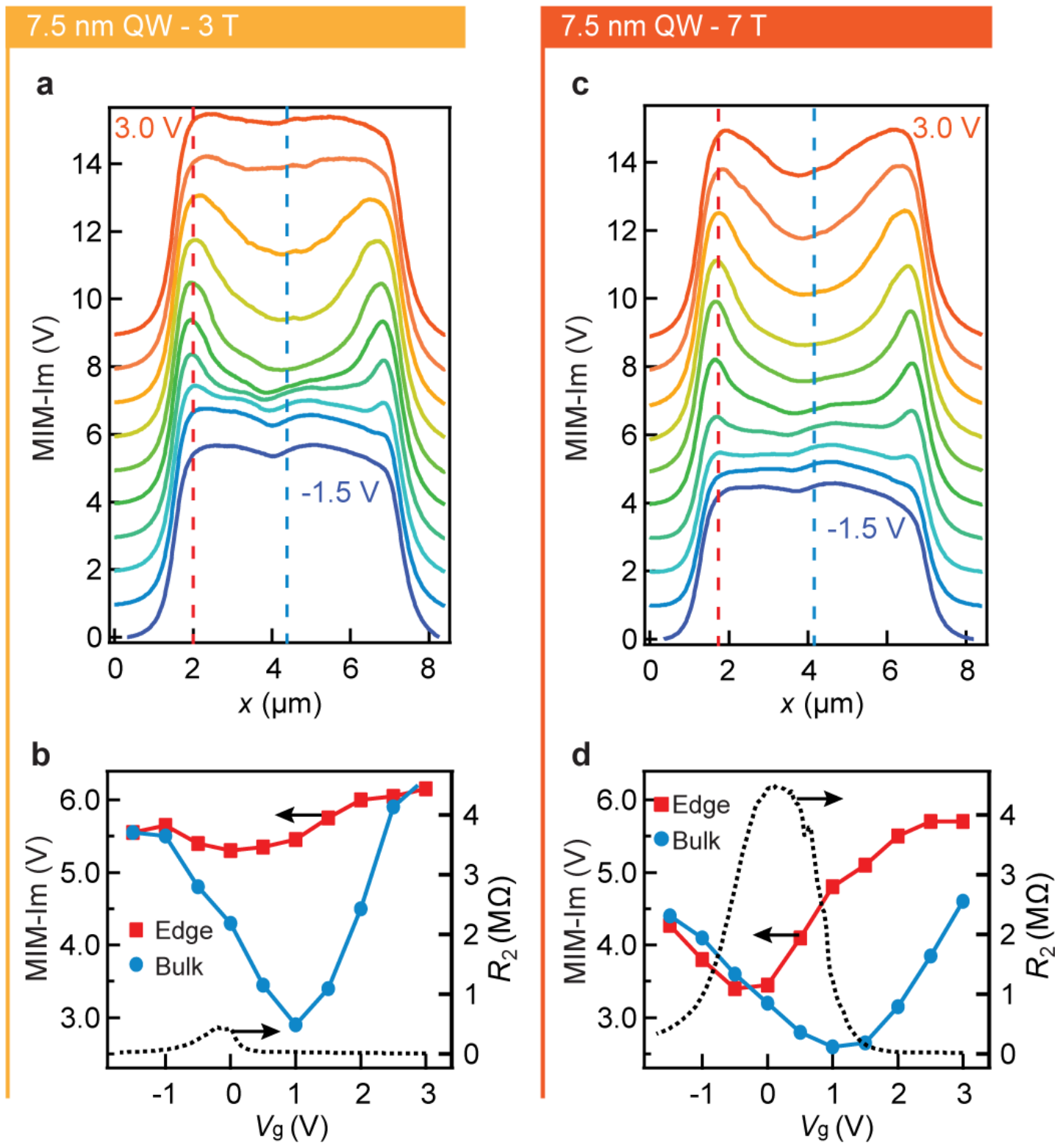

**Figure 3. Transport and MIM line-cut analysis for the 7.5 nm device at 3 T and 7 T. (a)** Averaged line-cuts of the center ⅓ section of the real space MIM-Im images in the 3 T column of Fig. 2j, with a 1.0 V offset in MIM signal between each gate voltage. Gate voltage is from -1.5 V to 3.0 V in 0.5 V step. **(b)** Edge and bulk MIM-Im signal (at the red and blue dashed lines in **a**) and 2-terminal resistance plotted against gate voltage. **(c)(d)** Same plots for the 7 T column. Transport is clearly dominated by edge conduction in the bulk gapped regime in both fields, confirming the extended nature of the edge conduction observed by MIM.